\documentclass[lenghtcheck,twocolumn,subeqn,amsmath,floats,tabularx,unsortedaddress,superscriptaddress,nofootinbib,prl,amssymb]{revtex4}
\usepackage{graphicx}
\usepackage{tabularx}  

\begin{document}
\title{Phonon momentum and damping of mechanical resonators}
\author{A. Borrielli}
\affiliation{Institute of Materials for Electronics and Magnetism,
Nanoscience-Trento-FBK Division,
 38123 Povo, Trento, Italy}
\affiliation{INFN, Gruppo Collegato di Trento, Sezione di Padova, 38123 Povo, Trento, Italy}

\author{E. Serra}
\affiliation{Interdisciplinary Laboratory for Computational Science (LISC), FBK-University of Trento, 38123 Povo (Trento), Italy}
\affiliation{INFN, Gruppo Collegato di Trento, Sezione di Padova, 38123 Povo, Trento, Italy}

\author{L. Conti}
\affiliation{INFN, Sezione di Padova, Via Marzolo 8, 35131 Padova, Italy}
 
\author{M. Bonaldi}
\email[Electronic mail: ]{bonaldi@science.unitn.it}
\affiliation{Institute of Materials for Electronics and Magnetism,
Nanoscience-Trento-FBK Division,
 38123 Povo, Trento, Italy}
\affiliation{INFN, Gruppo Collegato di Trento, Sezione di Padova, 38123 Povo, Trento, Italy}

\begin{abstract}
The concept of physical momentum associated to phonons in a crystal, complemented with some fundamental reasoning, implies measurable effects in crystals even at a macroscopic scale.   We show that, in close analogy with the transfer of momentum in the kinetic theory of gases, physical momentum carried by of phonons couples the thermal and the velocity field in a vibrating crystal. Therefore an heat flow applied to a vibrating crystal can sustain or damp the oscillation, depending on the interplay between the temperature and the velocity gradient. We derive the general equations of this effect and show that its experimental confirmation is  within reach of current technology.
\end{abstract}
\maketitle

In a crystalline solid, a phonon of wave vector $\textbf{k}$ carries a "crystal momentum" $\hbar\textbf{k}$ that participates in the interaction with particles or other elementary excitations that similarly carry their own crystal momentum. In fact  the ions of the crystal can be seen as scattering centers, and  the waves emitted from each ion have constructive interference  only if the total crystal momentum is conserved up to an arbitrary reciprocal lattice vector.
Therefore crystal momentum is not related to a physical momentum eventually associated to phonons, as they describe the relative internal motion of the atoms, and are not associated with any linear momentum of the crystal as a whole. 
The limits of this description becomes evident when the  translational invariance of the crystal is  broken, as  in a finite-size body, or in case of randomly distributed scattering centers.
In this case the anharmonic terms couple the relative internal motion with the center of mass motion, and phonons carry both crystal momentum $\hbar \textbf{k}$ and the physical linear momentum $\textbf{p}=3 \gamma \hbar \textbf{k}$ \cite{sorbello,faber99,lee06}. The connection between these two forms of momentum is a measure of the anharmonicity and is given by the Gr$\ddot{\textrm{u}}$neisen parameter $\gamma$ of the lattice, which measures  the relation between the volume of a crystal lattice and its vibrational properties. 

The intimate connection between phonon momentum and thermal expansion has been first pointed out by  Brillouin \cite{brillouin}, who showed that thermal expansion can be viewed as the reaction of the crystal lattice to the internal pressure due to phonons. More recently, an experiment has demonstrated that the collision of phonons can move a small cargo along a carbon nanotube \cite{barreiro08}. The concept of corpuscular phonon allows also a theoretical description of some dynamical phenomena in thermoelasticity, such as dynamical thermal expansion in thin optical windows \cite{lee09}, and is at the basis of the phonon pumping mechanism recently proposed for the cooling of nanodevices \cite{chamon11}.
Here we show that phonon momentum transfer couples the thermal and the velocity field in a vibrating crystal and we predict measurable effects in crystals even at a macroscopic scale.

This issue is relevant for a thorough understanding of oscillating systems of every size and for setting the limits to the performances of many instruments and devices. In fact resonant mechanical systems serve as a basis for a number of sensors and fundamental studies \cite{aguiar,favero09,poggio10,hough11}. Their size covers orders of magnitude, from micrometers to some meters, and their motion is detected, by electrical or optical readouts, with high sensitivity, sometimes limited only by the uncertainty principle. In these instruments  a low damping of the resonant motion is often crucial to achieve the desired performances, as in the case of  gravitational wave detectors \cite{saulson} or micro-mechanical resonators \cite{poot}. Here the laser beam used in optical readout usually heats some parts of the resonator: thermal gradients up to 10 K are expected in some components of the future cryogenic gravitational wave detectors \cite{puppo11}, while gradients of about 1 K are currently applied in cryogenic micro-resonators with optical readout \cite{poggio10}.

As discussed above, a physical linear momentum  $\textbf{p}=3 \gamma_{\textbf{k}s} \hbar k\,\textbf{i}_\textbf{k}$ is associated to a phonon of the branch $s$, where $\textbf{i}_\textbf{k}$ determines the direction of  propagation of the phonon and $\gamma_{\textbf{k}s}$ is the Gr$\ddot{\textrm{u}}$neisen parameter of the normal mode. As long as we deal with crystals at temperatures much smaller than the Debye temperature (for instance $\Theta_D=645 \;$K for silicon), we  can use the simple dispersion relation $\omega_s(\textbf{k})=c\, k$ and write $\textbf{p}=3 \gamma_{\textbf{k}s} \hbar \frac{\omega_k}{c}\textbf{i}_\textbf{k}$. Accordingly,  an equivalent mass $m_k=3 \gamma_{\textbf{k}s} \hbar \frac{\omega_k}{c^2}$,  arising from the coupling of the oscillatory motion of the atoms with the center of mass of the crystal, is associated to a phonon moving with velocity $c\,\textbf{i}_\textbf{k}$.  
In case of a phonon in a moving crystal, this relation must be complemented with a term depending by the motion of the center of mass. 
If we push forward the analogy between a phonon and a particle, it is natural to assign  a momentum:
\begin{equation}
\label{eq:momentumrelative_main}
\textbf{p}=3  \gamma_{\textbf{k}s} \hbar  \frac{\omega_k}{c^2} \left( c\,\textbf{i}_\textit{k} + \textbf{v}_Q \right)
\end{equation}
if the phonon is in a crystal moving with velocity $\textbf{v}_Q$ in respect to an observer at rest. 
This relation transforms the physical momentum between reference systems in relative motion, considering that the velocity of the phonon relative to the crystal is constant.
If the phonon is traveling in a crystal having parts moving with different velocities, we expect that it will couple in sequence with the parts of the crystal crossed by its trajectory, changing its frequency and/or direction as it moves. As a result of this interaction, the phonon will actually transfer momentum across the crystal. Even if this issue deserve dedicated molecular dynamics studies, as a first approach 
we  assume  that the phonon transfers to the crystal its energy  and momentum when suffering a collision, starting a new path with the frequency and momentum corresponding to the local temperature and velocity.  

\begin{figure}[t!]
\includegraphics[width=8.6cm,height=67mm]{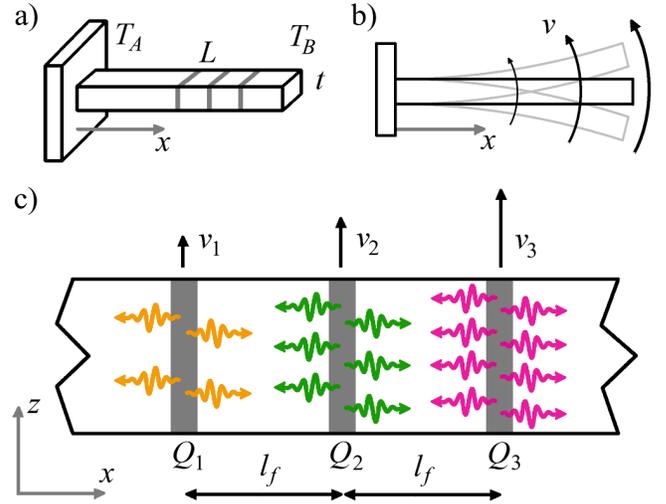}
\caption{(Color online) a) Heated cantilever beam, of length $L$ and thickness $t$. The temperature increases along the $x-$axis from $T_A$ to $T_B$. Three thin sections $Q_k$, at intervals of length $l_f$, are shown in gray. b) The cantilever oscillates in its first bending resonance mode. When it goes through a zero crossing, the local velocity is directed along the $z-$axis and its amplitude increases along the $x-$axis.  c) Detail of sections $Q_k$. The density and the frequency of the phonons increase from left to right, proportionally to the local temperature. As phonons travel for a distance $l_f$,  one half of the phonons starting from $Q_1$ and $Q_3$ reach the central section $Q_2$.
 \label{fig:cantilever}}
\end{figure}

To understand the physical meaning of these assumptions we propose a simplified model in Figure \ref{fig:cantilever}. An oscillating cantilever beam is heated from its free end, so that both the temperature and the amplitude of the velocity increase along the $x-$axis. We consider three thin sections of thickness $\Delta x\ll l_f$, say $Q_j$ with $j=1-3$, at intervals of length $l_f$ along the cantilever, with $l_f$ the Mean Free Path (MFP) of the phonons. Due to the temperature dependence of the phonon density, a larger number of  phonons and with higher energy are emitted from the section $Q_3$  in respect to those emitted from $Q_1$. Following the dominant phonon approximation \cite{pohl70}, we assume that the whole thermal energy in a section can be ascribed to  phonons of a single frequency $\bar{\omega}$. Then, in the time $\mathrm{d}t$, we have $(\bar{n}_1,\bar{n}_2 ,\bar{n}_3)$ phonons with frequency $(\bar{\omega}_1,\bar{\omega}_2,\bar{\omega}_3)$ starting respectively from    ($Q_1,\,Q_2,\,Q_3$), with $\bar{n}_3>\bar{n}_2>\bar{n}_1 $ and $\bar{\omega}_3>\bar{\omega}_2>\bar{\omega}_1$ (Figure \ref{fig:cantilever}c).
Now we consider the balance of the momentum  $p_z$ directed along the $z-$axis, at the central section $Q_2$ when the cantilever goes through a zero crossing. We assume that all phonons starting from  $Q_1$ and $Q_3$, travel for a length $l_f$ and then suffer a collision with the lattice in $Q_2$. According to Eq. \eqref{eq:momentumrelative_main}, phonons carry a momentum proportional to $\bar{\omega}_j  v_j $, then: 
\begin{equation}
\frac{\mathrm{d} p_z}{\mathrm{d}t} \propto   \frac{1}{2}\bar{n}_1\bar{\omega}_1 v_1+\frac{1}{2}\bar{n}_3\bar{\omega}_3 v_3-\bar{n}_2\bar{\omega}_2 v_2 
\end{equation}
where we take in account that only one half of the phonons starting in the $Q_1,Q_3$ are emitted towards $Q_2$. It is evident that, in a linear approximation of the velocity $v_2=(v_1+v_3)/2$, we have $p_z=0$ in the isothermal case, when phonon density and frequencies do not changes along the cantilever. In this case the excess momentum coming from $Q_3$ is compensated by a smaller momentum from $Q_2$. On the contrary a thermal gradient 
changes the number and the frequency of phonons coming from $Q_1$ and $Q_2$ and we easily obtain:
\begin{equation}
\label{eq:simpleq}
\frac{\mathrm{d} p_z}{\mathrm{d}t}  \propto   \frac{ \partial(\bar{n} \bar{\omega} )}{\partial x} \frac{ \partial v  }{\partial x} l_f^2  
\end{equation}
showing that within this framework a combination of thermal and velocity gradients can transfer physical momentum along the cantilever. 

To evaluate the possible implications of this effect on the dynamics of a resonator, we have to put Eq. \eqref{eq:simpleq} on a more solid ground, in close analogy with the theory of transfer of momentum in the kinetic theory of gases \cite{loeb}.
We consider the point $P\equiv (x,y,z)$ in a crystal, 
where the temperature  is given by a  scalar field $T(x,y,z)$ and the velocity by a vector field $\textbf{v}(x,y,z)$. 
We assume that a steady state has been reached  and evaluate the balance of momentum in a small sphere $\Lambda$ of radius $\mathrm{d}\lambda$ centered at $P$. We place a polar coordinate system at $P$, and consider a small volume $\mathrm{d}\sigma=r^2\sin\theta\,  \mathrm{d} r\, \mathrm{d}\theta \,\mathrm{d}\phi$ at the point $\textbf{r}\equiv (r,\theta,\phi)$, say $Q$, in this coordinate system  (see Fig. 1 of the supplementary material \cite{supplementary}). Both volumes are small in the sense that their characteristic size is much smaller than the MFP  $l_f$. For each phonon branch $s$, in $\mathrm{d}\sigma$ are $n_s(T_Q, \textbf{k})\mathrm{d}\sigma$ phonons with wave vector $\textbf{k}$, where $T_Q$ is the value of the temperature at the point $Q$. Each phonon starts a free path $c/l_f$ times per second, where $c$ is the group velocity, therefore  a number of $\frac{n_s(T_Q,\textbf{k})c}{l_f}\mathrm{d}\sigma \,\mathrm{d}t$ phonons will leave the volume $\mathrm{d}\sigma$ in the time $\mathrm{d}t$. 
Of these, the number in a cone of base $\pi (\mathrm{d}\lambda)^2$ will have paths directed towards the small sphere if all directions are equally probable. That is, $\frac{\pi (\mathrm{d}\lambda)^2}{4 \pi r^2}n_s(T_Q,\textbf{k})\frac{c}{l_f}\mathrm{d}\sigma\,\mathrm{d}t$ phonons will leave  headed for the sphere along $\textbf{r}$. 
Of these, a fraction only will suffer a collision within the sphere $\Lambda$: if we consider for instance only phonons directed exactly along $-\textbf{r}$,  the fraction $e^{-(r- \mathrm{d}\lambda )/l_f}$  will succeed in reaching the sphere at $P$, while the fraction $e^{-(r+\mathrm{d}\lambda )/l_f}$ will cross it without impact. As a result the number of collisions  within  $\Lambda$ is   $\frac{2}{3}e^{-r/l_f}\frac{2 \mathrm{d}\lambda}{l_f}$ , where the factor $\frac{2}{3}$ accounts for the reduction of the length of the path on the borders of the sphere. Hence the number of phonons with wave vector $\textbf{k}$ starting in $d\sigma$ and colliding within the sphere $\Lambda$ is:
\begin{equation}
\label{eq:totalnumberSphere}
\mathrm{d}N=\frac{4}{3} {\pi (\mathrm{d}\lambda)^3} n_s(T_Q,\textbf{k})\frac{c}{4 \pi r^2l_f^2}e^{-r/l_f}\mathrm{d}\sigma \,\mathrm{d}t   
\end{equation}
The first term is the volume of the sphere $\Lambda$, that we approximate as a cubic volume element $\mathrm{d}V=\mathrm{d}x\, \mathrm{d}y\, \mathrm{d}z$ to allow the integration over the crystal \cite{approx}. The collision rate at $P$ is then:
\begin{equation}
\label{eq:totalnumberCube_main}
\frac{\mathrm{d}N}{\mathrm{d}t}=  \,n_s(T_Q,\textbf{k})\frac{c}{4 \pi r^2 l_f^2}\,e^{-r/l_f} \,\mathrm{d}\sigma \, \mathrm{d}V   
\end{equation}
The components of the total rate of the momentum entering the element $dV$  are obtained by integrating the product of equation \eqref{eq:totalnumberCube_main} and \eqref{eq:momentumrelative_main}. Here $\textbf{k}$ varies over the full phonon spectrum and the volume $\mathrm{d} \sigma$, placed at the source point $Q$, over the volume of the body $V_B$: 
\begin{eqnarray}
\label{eq:entering_dt_main}
\frac{\mathrm{d}\textbf{p}}{\mathrm{d}t}&=& \mathrm{d}V  \int_{V_B} \mathrm{d} \sigma \left[  \sum_{\textbf{k}s}  \,\gamma_{\textbf{k}s}\,\hbar\,\omega_k\, n_s(T_Q,\textbf{k})\right]  \times \nonumber\\  && \times   \frac{3    }{4 \pi r^2   l_f^2 c}\,e^{-r/l_f}  \left(-c\,\textbf{i}_r+\textbf{v}_Q  \right)  
\end{eqnarray} 
where the direction of propagation of the phonons has been approximated as $\textbf{i}_\textbf{k}\simeq -\textbf{i}_r$. This equation can be evaluated after some preliminary consideration. 
First, the volume integral can be extended over all space, because only a sphere of radius comparable with a few mean free path actually contributes to the volume integral, due to the exponential term $e^{-r/l_f}$. 
Second, we note that $\hbar\,\omega_k\, n_s(T,\textbf{k})=\int_0^T c_{vs}(\textbf{k})\mathrm{d}T$, where 
$c_{vs}(\textbf{k})$ is the contribution of the mode $(\textbf{k}\,s)$ to the specific heat  $C_v=\sum_{\textbf{k} s}c_{vs}(\textbf{k})$. Then the sum over the phonon spectrum  is $\sum_{\textbf{k}s}  \,\gamma_{\textbf{k}s}\int_0^T c_{vs}(\textbf{k})\mathrm{d}T=\int_0^T\gamma(T) C_v\mathrm{d}T$, where $\gamma(T)=\sum_{\textbf{k} s}\gamma_{\textbf{k}s}c_{vs}(\textbf{k})/C_v$ is the overall Gr$\ddot{\textrm{u}}$neisen parameter.
In general this integral has a complex functional dependence on the temperature \cite{supplementary}, as $\gamma(T)$ may change in a complicated way.  On the other hand at very low temperatures (below 10 K) we have $\int_0^{T_Q}\gamma(T) C_v \mathrm{d}T\simeq \gamma_0 \frac{B_v}{4} T_Q^4$, with $B_v$ determined by the relation $C_v\simeq B_v T^3$ and $\gamma_0=\lim_{T\rightarrow 0} \gamma(T)$. At intermediate temperatures (30-70 K for silicon) this relation remains a good approximation with $\gamma_0$ an effective Gr$\ddot{\textrm{u}}$neisen parameter \cite{supplementary}. We then expand  $\textbf{v}_{Q}$ and $T_Q^4$   to I order in $\textbf{r}$ around the point $P$, and obtain for the components of $\frac{\mathrm{d}\textbf{p} }{\mathrm{d}t}$:
\begin{eqnarray}
\label{eq:5terms}
 \frac{\mathrm{d}p_\alpha}{\mathrm{d}t}  &&\simeq \; \mathrm{d}V  \,  \gamma_0  \frac{B_v}{4}   \times\\ 
&&\times\left(  \frac{3 T_P^4 }{ l_f c }v_{P\alpha} -4T_P^3   \, \frac{\partial T_P}{\partial \alpha}+    \frac{8T_P^3l_f}{ c} \, \nabla T_P\cdot\nabla  {v}_{P\alpha}\right) \nonumber
\end{eqnarray}
where $\alpha =x,y,z$ and all derivatives are evaluated at the point $P$. The details of the calculation, with a discussion of II order terms not shown here, can be found in the supplementary material \cite{supplementary}.
  
To complete the momentum budget we must consider the rate of momentum loss due to phonons starting a new free path in the element $\mathrm{d}V$. For each direction $(\theta, \phi)$
 a number of $\mathrm{d}V\,n_s(T_P,\textbf{k})\frac{c}{l_f} \, \frac{\sin \theta \,d\theta \,d\phi}{4\pi}\,\mathrm{d}t$ phonons will leave the element in the time $\mathrm{d}t$, each carrying away the momentum given by Eq. \eqref{eq:momentumrelative_main}. It is straightforward to show that   this contribution, when integrated over the phonon spectrum and all directions, compensates the first term (thermal equilibrium term) of Eq. \eqref{eq:5terms}. 
 
The second term (phonon pressure term) is  proportional to the temperature gradient and describes the balance of momentum along the direction of propagation $\textbf{i}_\textbf{k}$. In fact, the distribution of phonons arriving from the hot part is shifted toward higher frequency, therefore the resultant momentum rate is directed against the temperature gradient. The force associated to this momentum rate is supported by the crystal through the elasticity modulus and induces a nonuniform expansion across the volume of the solid body. Hence our model extends to the case of a thermal gradient the well known connection between phonon momentum and thermal expansion \cite{brillouin,ashcroft}. 

The third term (damping term) is proportional to the gradients of the temperature and of the velocity fields, and can have striking effects in the case of an oscillating crystal with a thermal gradient. In fact the rate of momentum transferred from the phonon gas to the crystal is equivalent to a force $\textbf{f}_P=\frac{\mathrm{d}\textbf{p}}{\mathrm{d}t}$. If the point $P$ in the time $\mathrm{d}t$ changes its position by $\mathrm{d}\textbf{x}_P$, the work done by this force is $\textbf{f}_P\cdot \mathrm{d}\textbf{x}_P$. This work can be easily evaluated if the system is oscillating with angular frequency $\omega_0$. In this case the components of the displacement field are  $u_\alpha(x,y,z,t)=G_u  f_\alpha(x,y,z) \sin(\omega_0 t)$, where $G_u$ is the magnitude of the amplitude oscillation and $f_\alpha(x,y,z)$ the displacement function. The  velocity field and its gradient are respectively $v_\alpha(x,y,z,t)= \omega_0 G_u   f_\alpha(x,y,z) \cos (\omega_0 t)$ and     $\nabla  v _\alpha = \omega_0 G_u   \nabla  f_\alpha   \cos (\omega_0 t)$. If we consider that the force $\textbf{f}_P$ is much smaller than the elastic forces at the origin of the oscillatory behavior, we can safely assume that it does not alter in a significant manner the displacement field $u_\alpha(x,y,z,t)$. Then the work done in an oscillation cycle is simply $\delta L_P=\int_0^{{2\pi}/{\omega_0} }\textbf{f}_P\cdot \textbf{v}_P\, \mathrm{d}t$, and 
this local contribution can be integrated over the volume of the crystal to obtain the total work:
\begin{equation}
\label{eq:work_main}
L_{Ph}= \frac{ 2 \pi   \gamma_0  \,  l_f  B_v }{c} \omega_0 G_u ^2\; {\sum_\alpha\int_V \left( T^3\, (\nabla T\cdot\nabla  {f}_{\alpha})\,{f}_{\alpha}
\right)\mathrm{d}V} 
\end{equation}
This equation is the main result of this paper and must be carefully analyzed to understand the possible implications on the dynamics of the resonator.

We note first that the work $L_{Ph}$ is proportional to the squared amplitude of the oscillation, just like the energy losses due to internal friction in the resonator. On the other hand it can be positive or negative, depending on the interplay between velocity and temperature gradients, therefore the superimposed heat flow can sustain or damp the oscillation of the body. 
To compare more easily this effect  with the usual dissipative phenomena in resonating bodies, we define the loss angle due the phonon momentum transport as $\phi_{ph}=-\frac{1}{2\pi}	\frac{L_{Ph}}{W}$, where $W=\frac{1}{2}\rho\, G_u^2 \omega_0^2 \sum_\alpha \int_V f_\alpha^2 \mathrm{d}V$  is the total energy stored in the resonator. From Eq. \eqref{eq:work_main} we have:
\begin{equation}
\label{eq:damping_main}
\phi_{ph}= -\frac{ 2    \gamma_0     l_f  B_v }{\rho\omega_0 c }\;\frac{\sum_\alpha\int_V \left( T^3\, (\nabla T\cdot\nabla  {f}_{\alpha})\,{f}_{\alpha}
\right)\mathrm{d}V}{\sum_\alpha \int_V f_\alpha^2 \mathrm{d}V}
\end{equation}
where positive values of $\phi_{ph}$ mean a damping of the oscillation. 
Besides this phononic contribution, the energy loss in a resonator results from the contribution of several dissipation mechanism, that collectively determine the background dissipation $\Delta W_{bg}$ and the background loss angle $\phi_{bg}=\frac{1}{2\pi}	\frac{\Delta W_{bg}}{W}$. The most common dissipation mechanism are thermoelastic and structural loss in the resonator and  energy loss through the support. In this respect, we note that silicon resonators based on cantilevered beams have demonstrated a background loss as low as  $10^{-6}$ at cryogenic temperatures, with a reproducibility of about 20\% \cite{reid06}.

\begin{figure}[t!]
\includegraphics[width=8.6cm,height=50mm]{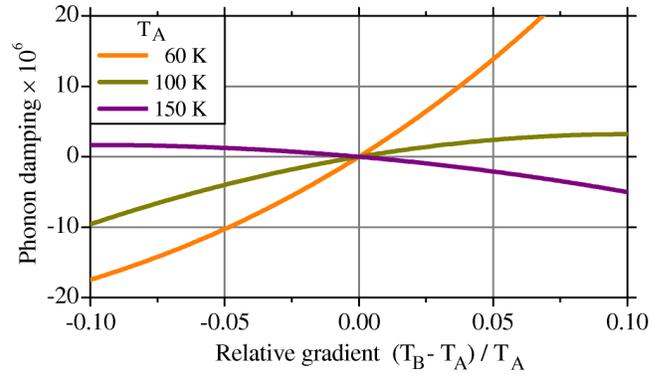}
\caption{(Color online) Expected damping due to the phonon momentum transport for the first bending resonance mode (55 hz) of the silicon cantilever. The curve at 60 K is obtained in the low temperature approximation (Eq. \eqref{eq:damping_main}) with $\gamma_0=-0.8$, while curves at 100 K and 150 K are evaluated by taking in account the temperature dependence of the   Gr$\ddot{\textrm{u}}$neisen parameter \cite{supplementary}. 
A positive thermal gradient can be obtained by heating the free end of the beam. Then, depending on the base temperature, the additional loss  $\phi_{ph}$ can augment or reduce the background damping of the resonator.   Typical background loss for a silicon cantilever of this size is $\phi_{bg}\simeq 10^{-6}$.
 \label{fig:results}}
\end{figure}

For the cantilever of Fig. \ref{fig:cantilever} the temperature profile is  $T(x)=(T_B-T_A)x/L +T_A$ and the displacement is fully described by the the component $f_z$. We use the variable  $\xi=x/L$, $\xi \in(0,1)$, and write the modal shapes  as:
\begin{eqnarray}
\label{eq:cantilever_shape}
&&f_z(\xi)= \left[\sin(\beta ) - \sinh(\beta)\right]   \left[\sin(\beta \xi) - \sinh(\beta \xi)\right] +\nonumber\\
    && + \left[\cos(\beta) +  \cosh(\beta) \right] \left[\cos(\beta\xi) - \cosh(\beta \xi) \right]
\end{eqnarray}
where  $\beta=1.875$ for the first resonant mode \cite{meirovitch}. The frequency of the oscillation is obtained as $\omega^2=\frac{\beta^2 t}{L^2}\sqrt{\frac{Y}{12 \rho}}$, where $Y$ and $\rho$ are respectively Young modulus and density of the crystal.  In Fig. \ref{fig:results} we show the expected damping for the lowest frequency mode of a silicon cantilever ($Y=169\,$GPa, $\rho=2329\,$kg/m$^3$) of thickness 100$\,\mu$m and lenght 50$\,$mm. We use  the average  MFP  from measurements of thermal conductivity accumulation distribution \cite{minnich11}, and $\gamma(T)$ from thermoelastic measurements \cite{sparks67,gauster71}. In this case Eq. \eqref{eq:damping_main} cannot be used below 60 K, as the MFP would become larger than the thickness of the cantilever and boundary scattering dominate the transport.

Now we can trace out the main features of the theory and evaluate its experimental relevance.
At first we note that, with a relative gradient of 10\%, the magnitude of $\phi_{ph}$ can be well above the typical background loss in this systems \cite{reid06}. Second, according to the sign of $\gamma(T)$ and of the gradient, the additional loss  $\phi_{ph}$ can augment or reduce the overall damping of the resonator. 
Therefore with the application of a thermal gradient it is possible to reach a regime where the dynamics of the resonator is dominated by the phononic damping. In this case, the  loss angle can be much higher than expected background, spoiling off the performances of the device, or could be considerably reduced, possibly driving the system towards region of dynamic instability. From another perspective this effect could allow the design of micro-resonators exploiting the transport of phonon momentum as a power source, drawing energy from the thermal gradient to sustain their oscillatory motion.  
We point out that phonon momentum effects turn out to be accessible in a macroscopic system  thanks to the extreme energy sensitivity of low loss resonant systems, while a wider variety of mechanical effects can be observed at the nanoscale \cite{barreiro08}.  

We have extended the phonon momentum concept to the case of an oscillating crystal, and within this framework we have evaluated the dynamic behavior of an arbitrarily shaped resonator with an applied thermal gradient. In case of a cantilever we have found a number of features that are at experimental reach and could have a significant role in the performances of future devices for precision measurements. For this reason we hope that these ideas provoke new experimental and theoretical efforts aimed toward understanding the effect of phonon momentum on dissipative processes in resonators.

We acknowledge the contribution of the European Research Council under the European Community's Seventh Framework Programme (FP7/2007-2013)/ERC grant agreement no. 202680.


\begin{thebibliography}{99}
\bibitem{sorbello}R. Sorbello, Phys. Rev. B \textbf{6}, 4757 (1972).
\bibitem{faber99}T.E. Faber, Phil. Mag. B \textbf{79}, 1445 (1999).
\bibitem{lee06}Y.C. Lee and  W.Z. Lee, Phys. Rev. B \textbf{74} 172303 (2006).
\bibitem{brillouin}L. Brillouin, \textit{Les tenseurs en mecanique et en elasticite} (Masson, Paris, 1938).
\bibitem{barreiro08}A. Barreiro, R. Rurali, E.R. Hern\'andez, J. Moser, T. Pichler, L. Forr\'o, A. Bachtold, Science   \textbf{320}  775 (2008).
\bibitem{lee09}Y.C. Lee, J. Phys.: Condens. Matter \textbf{21} 325702 (2009).
\bibitem{chamon11}C. Chamon, E.R. Mucciolo, L. Arrachea and R.B. Capaz,  Phys. Rev. Lett. \textbf{106}, 135504 (2011).
\bibitem{aguiar} O. D. Aguiar, Res. Astron. Astrophys. \textbf{11},  1 (2011).
\bibitem{hough11}M. Pitkin, S. Reid,  S. Rowan and J. Hough, Living Rev. Relativity \textbf{14}, URL (cited on 2012-01-5): http://www.livingreviews.org/lrr-2011-5 (2011).
\bibitem{favero09}I. Favero and K. Karrai,   Nature Photonics \textbf{3} 201 (2009).
\bibitem{poggio10}M. Poggio and C. L. Degen, Nanotechnology 21, 342001 (2010).
\bibitem{saulson}P. R. Saulson, Phys. Rev. D \textbf{42}, 2437 (1990).
\bibitem{poot} M. Poot and H.S.J. van der Zant, Physics Reports \textbf{511}, 273 (2012).
\bibitem{puppo11}P. Puppo and F. Ricci,  General Relativity and Gravitation \textbf{43}, 657 (2011).
\bibitem{pohl70}V. Narayanamurti  and R. O. Pohl,  Rev. Mod. Phys. \textbf{42}, 201  (1970). 
\bibitem{loeb}L.B. Loeb, \textit{The Kinetic Theory of Gases} (McGraw-Hill, New York, 1934).
\bibitem{approx}From the comparison of the volume of a cube inscribed and circumscribed about a sphere we estimate the approximation error as $\pm 20\%$.
\bibitem{supplementary}See supplementary material at http://link.aps.org/
\bibitem{ashcroft}N. W. Aschcroft and N. D. Mermin, \textit{Solid State Physics} (Holt, Rinehart, and Winston, New York, 1976).
\bibitem{meirovitch}L. Meirovitch, \textit{Principles and Techniques of Vibrations}, (Prentice-Hall International, London, 1996).
\bibitem{minnich11}A. J. Minnich, J. A. Johnson, A. J. Schmidt, K. Esfarjani, M. S. Dresselhaus, K. A. Nelson, and G. Chen, Phys. Rev. Lett. 107, 095901 (2011). 
\bibitem{sparks67}P.W. Sparks and C.A. Swenson, Phys. Rev.   \textbf{163}, 779 (1967).
\bibitem{gauster71}W.B. Gauster, Phys. Rev. B \textbf{4}, 1288 (1971).
\bibitem{reid06}S. Reid, G. Cagnoli, D.R.M. Crooks, J. Hough, P.Murray, S.Rowan, M.M. Fejer,
R. Route, S. Zappe, Physics Letters A \textbf{351},  205  (2006).

\end{thebibliography}
\end{document}


\begin{center}
{\Large  \textbf{Supplemental material to:}}
\end{center}
\begin{center}
{\Large  \textbf{"Phonon momentum and damping of mechanical resonators"}} 
\end{center}
\begin{center}{\Large
A. Borrielli, E. Serra, L. Conti and M. Bonaldi}
\end{center}
\begin{center}
{\Large
email: \texttt{bonaldi@science.unitn.it}}
\end{center}

\vspace{1cm}

{\large
This PDF file includes the following material:
\begin{itemize}
\item Coordinate system (Figure 1)

\item Temperature integral: approximation for silicon (Figure 2)

\item Volume integral: first and second order approximation

\item Volume integral: medium temperature approximation for silicon
\end{itemize}}
\newpage

\section{Coordinate system}
\begin{figure}[h!]
\includegraphics[width=8.6cm,height=93mm]{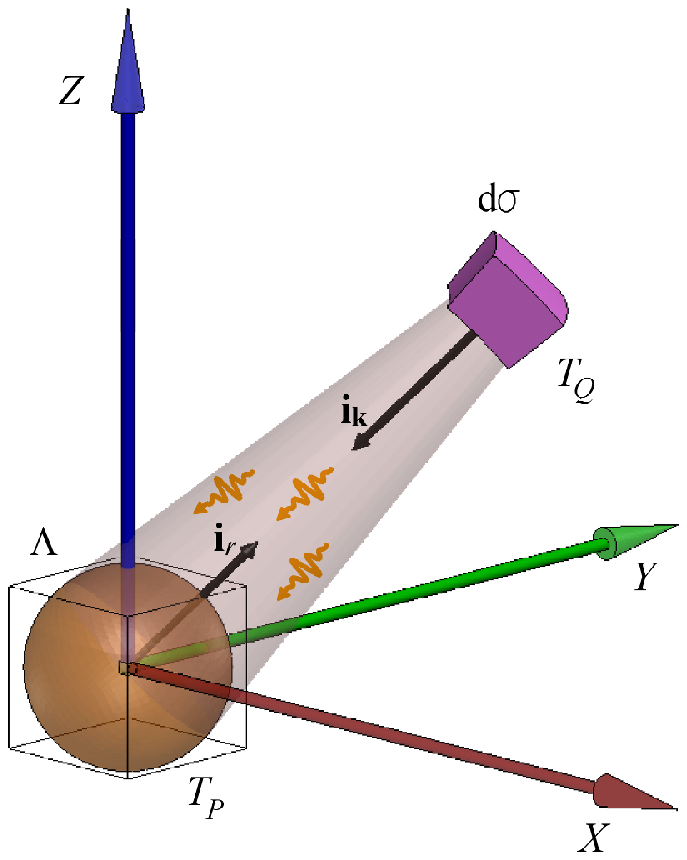}
\caption{Coordinate system used in the volume integration.
 \label{fig:volume_integration}}
\end{figure}
\newpage  
  
\section{Temperature integral: approximation for silicon} 
The sum over the phonon spectrum in Equation (6) of the main text reduces to:
\begin{equation}
F(T_Q)\equiv \sum_{\textbf{k}s}  \,\gamma_{\textbf{k}s}\int_0^{T_Q}  c_{vs}(\textbf{k})=\int_0^{T_Q} \gamma(T) C_v  \mathrm{d}T
\end{equation}
where $\gamma(T)=\sum_{k s}\gamma_{\textbf{k}s}c_{vs}(\textbf{k})/C_v$ is the overall Gr$\ddot{\textrm{u}}$neisen parameter and may change in a complicated way with the temperature. For instance in silicon $\gamma(T)$ takes negative values between 19 K and 124 K.
On the other hand the strong temperature dependence of the specific heat $C_v$ can make negligible the low temperatures contribution  to the integral.

\begin{figure}[h!]
\includegraphics[width=8.6cm,height=94mm]{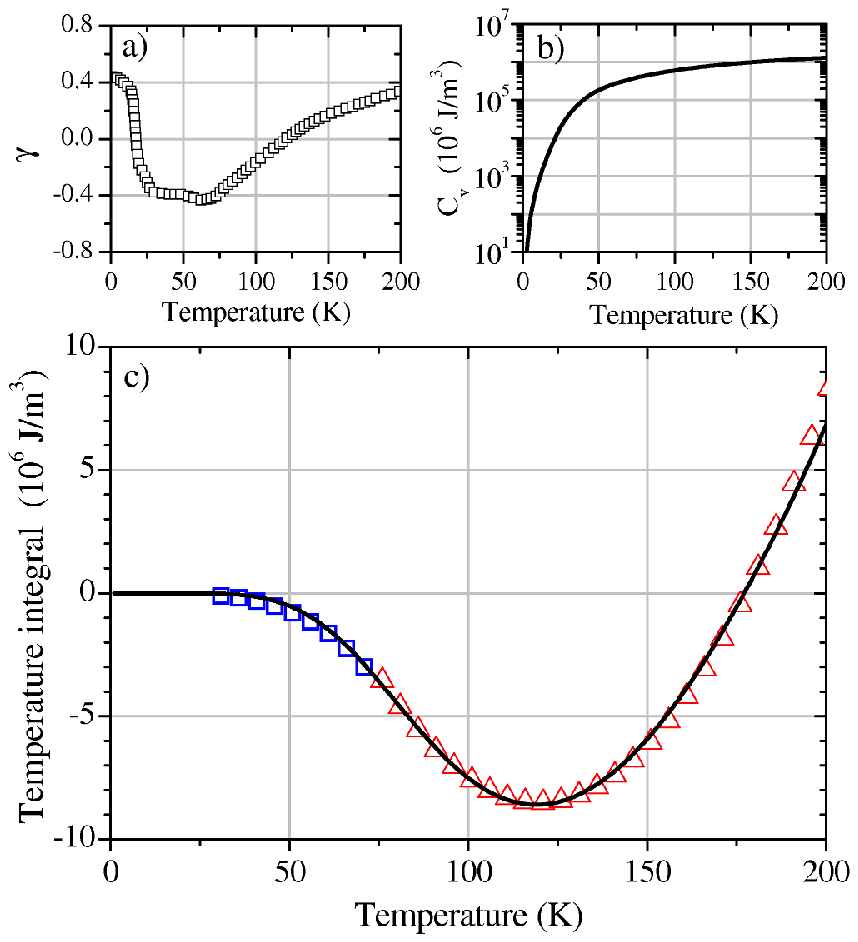}
\caption{Evaluation of the temperature integral $F(T)=\int_0^{T_Q} \gamma(T)\, C_v dT$ for the case of silicon. a) Temperature dependence of the  overall Gr$\ddot{\textrm{u}}$neisen parameter $\gamma(T)$ \cite{gauster71}; note that $\gamma(T)$ is negative from 18 K to 124 K, where silicon has a negative coefficient of thermal expansion. b) Specific heat $C_v$ of silicon \cite{mpdb}. c) Solid line: numerical evaluation of  $F(T)$. Hollow squares: points from the low temperature approximation Eq. \eqref{eq:FVLT} but with an effective Gr$\ddot{\textrm{u}}$neisen parameter $\gamma_0\simeq - 0.8$. Hollow triangles: points from the medium  temperature quadratic approximation Eq. \eqref{eq:FMT}.
 \label{fig:gruneisen}}
\end{figure}

In the case of silicon at very low temperatures (below 10 K) we have:
\begin{equation}
\label{eq:FVLT}
F_{VLT}(T_Q)=\simeq \gamma_0 \frac{B_v}{4} T_Q^4
\end{equation}
with $B_v=0.591\;$J/(m$^3$K$^4$) determined by the relation $C_v\simeq B_v T^3$ and $\gamma_0\simeq 0.4$ \cite{sparks67}. 
As shown in Figure \ref{fig:gruneisen}c, this relation remains a good approximation in the range (30-70 K) if we use an effective Gr$\ddot{\textrm{u}}$neisen parameter $\gamma_0\simeq 0.8$. 

In the medium temperature range (70-200) K, the temperature dependence of $\gamma(T)$ changes the functional form of $F(T)$, that now is well approximated by a quadratic functions:
\begin{equation}
\label{eq:FMT}
F_{MT}(T_Q)=F_0+ F_1 T_Q+ F_2 T_Q^2
\end{equation}
with $F_0=2.85\times 10^7$, $F_1=-6.17 \times 10^5$ and $F_2=2.57\times 10^3$.

\newpage
\section{Volume integral: first and second order approximation}

From Equation (6) of the main text, after the average over the phonon modes,
 we write the rate of momentum entering the volume element $dV$ as:
\begin{equation}
\label{eq:sup_entering_dt_main0}
\frac{\mathrm{d}\textbf{p}}{\mathrm{d}t}=\mathrm{d}V\,  \frac{3   }{4 \pi \, l_f^2 c }  \int_{V_B} \mathrm{d} \sigma F(T_Q)    \frac{1  }{  r^2  }\,e^{-r/l_f}  \left(-c\,\textbf{i}_r+\textbf{v}_Q  \right)  
\end{equation}
In the very low temperature case we have $F(T_Q)=F_{VLT}(T_Q)=\simeq \gamma_0 \frac{B_v}{4} T_Q^4$, with $\gamma_0=\lim_{T\rightarrow 0} \gamma(T)$. As shown in the Figure \ref{fig:gruneisen}, this relation remains valid at higher temperatures with $\gamma_0$ an effective Gr$\ddot{\textrm{u}}$neisen parameter. Therefore  Equation \eqref{eq:sup_entering_dt_main0} becomes:
\begin{equation}
\label{eq:sup_entering_dt_main}
\frac{\mathrm{d}\textbf{p}}{\mathrm{d}t}=\frac{B_v}{4} \frac{3  \gamma_0  }{4 \pi   l_f^2 c }\,\mathrm{d}V  \int_{V_B} \mathrm{d} \sigma \, T_Q^4     \frac{1  }{  r^2  }\,e^{-r/l_f}  \left(-c_i\textbf{i}_r+\textbf{v}_Q  \right)  
\end{equation}

With reference to the coordinate system shown in Figure \ref{fig:volume_integration}, we expand  $\textbf{v}_{Q}$ and $T_Q^4$ up to II order in $\textbf{r}$:
\begin{eqnarray}
\label{eq:vrII} 
{v}_{Q\alpha}&=& {v}_{P\alpha}+\textbf{r}\cdot \nabla  {v}_{P\alpha}+\frac{1}{2}\left( \textbf{r} \cdot \nabla  \right)^2{v}_{P\alpha}\\
\label{eq:TSII}
 T_Q^4&=&T_P^4+ \textbf{r}\cdot \nabla T_P^4+ \frac{1}{2}\,\left( \textbf{r} \cdot \nabla \right)^2{T}_{P}^4
\end{eqnarray}
where the derivatives are evaluated at the point $P$, corresponding to $\textbf{r}=0$.
The volume element is $d\sigma = r^2\sin\theta \mathrm{d}r \mathrm{d}\theta \mathrm{d}\phi$, therefore the volume integral for the component $\alpha$, with $\alpha=x,\,y,\,z$, is: 
\begin{eqnarray}
 \int_0^\infty \mathrm{d}r   \int_{0}^{\pi} \mathrm{d}\theta  \int_0^{2\pi}\mathrm{d}\phi \,\sin\theta \, e^{-r/l_f}   
\left(T_P^4+ \textbf{r}\cdot \nabla T_P^4+ \frac{1}{2} \left( \textbf{r} \cdot \nabla \right)^2{T}_{P}^4\right)
 \left( -c\,\textbf{i}_r\cdot \textbf{i}_\alpha+ {v}_{P\alpha}+\textbf{r}\cdot \nabla  {v}_{P\alpha}+\frac{1}{2}\left(\textbf{r} \cdot \nabla \right)^2{v}_{P\alpha} \right)
\end{eqnarray} 
where we have extended the volume integral over all space, because only a sphere of radius comparable with a few mean free path actually contributes to the volume integral, due to the exponential term $e^{-r/l_f}$. Here $\textbf{i}_r=\sin \theta \,\cos\phi\,\textbf{i}_x + \sin \theta \,\sin\phi\,\textbf{i}_y+ \cos \theta \,\textbf{i}_z$ and $\textbf{r}=r \textbf{i}_r$

We note that, as derivatives are evaluated at a fixed point,  each term is the projection of a constant vector along $\textbf{i}_r$. Then, in the angular average:
\begin{itemize}
\item  a single projection averages to zero due to the angular oscillating term:
\begin{eqnarray}
\label{eq:integralUno}
&& \int_0^{2\pi}\mathrm{d}\phi  \int_0^{\pi}\sin\theta\, \mathrm{d}\theta \left[ \textbf{i}_r\cdot \textbf{i}_\alpha \right] =0   \\
&&\int_0^{2\pi}\mathrm{d}\phi  \int_0^{\pi}\sin\theta\, \mathrm{d}\theta \left[ \textbf{i}_r\cdot \nabla  f_1(x,y,z) \right] = 0
\end{eqnarray}
where $f_1(x,y,z)$ is a scalar field and the gradient is evaluated at $\textbf{r}=0$.
\item The product of two projection gives a finite value: 
\begin{eqnarray}
\label{eq:relation1}
&&\int_0^{2\pi}\mathrm{d}\phi  \int_0^{\pi}\sin\theta\,\mathrm{d}\theta \left[ \textbf{i}_r\cdot \textbf{i}_\alpha\right] \left[ \textbf{i}_r\cdot \nabla f_1(x,y,z)\right]  =\frac{4}{3}\pi\,  \frac{\partial f_1(x,y,z)}{\partial \alpha}\\
\label{eq:relation2}
&&\int_0^{2\pi}\mathrm{d}\phi  \int_0^{\pi}\sin\theta\,\mathrm{d}\theta \left[ \textbf{i}_r\cdot \nabla f_1(x,y,z)\right] \left[ \textbf{i}_r\cdot \nabla  f_2(x,y,z)\right] =\frac{4}{3}\pi\,  \nabla f_1(x,y,z)\cdot\nabla  f_2(x,y,z)\\
&&\int_0^{2\pi}\mathrm{d}\phi  \int_0^{\pi}\sin\theta\, \mathrm{d}\theta  \left( \textbf{i}_r\cdot \nabla  \right)^2 f_1(x,y,z) = \frac{4}{3}\pi  \,\nabla^2 f_1(x,y,z)
\end{eqnarray} 
where $f_2(x,y,z)$ is another scalar field and again the gradients is evaluated at $\textbf{r}=0$.
\item  The product of three projection gives again a null value:
\begin{eqnarray}
  \int_0^{2\pi}\mathrm{d}\phi  \int_0^{\pi}\sin\theta\, \mathrm{d}\theta \left[\,\textbf{i}_r\cdot \textbf{i}_\alpha \right]\,\left[ \left( \textbf{r} \cdot \nabla  \right)^2 f_1(x,y,z) \right] =0
\end{eqnarray}  
\end{itemize}
As an example, we evaluate Equation \eqref{eq:relation1} in the case $\alpha=x$. Now $\left[ \textbf{i}_r\cdot \textbf{i}_\alpha\right]=\sin \theta \,\cos\phi$, therefore
\begin{eqnarray} 
&&  \int_0^{2\pi}\mathrm{d}\phi  \int_0^{\pi}\sin\theta\,\mathrm{d}\theta \left(  \sin \theta \cos\phi   \frac{\partial f_1}{\partial x}  + \sin \theta \sin\phi \frac{\partial f_1}{\partial y}+ \cos \theta  \frac{\partial f_1}{\partial z}  \right) \sin\theta\cos\phi=\nonumber\\
&&  \int_0^{2\pi}\mathrm{d}\phi  \int_0^{\pi}\sin\theta\,\mathrm{d}\theta \left(  \sin^2 \theta \cos^2\phi   \frac{\partial f_1}{\partial x}    \right)  =\nonumber\\
&& \frac{\partial f_1}{\partial x} \int_0^{2\pi}\cos^2\phi\, \mathrm{d}\phi  \int_0^{\pi}\sin^3\theta\,\mathrm{d}\theta = \frac{4}{3}  \,\pi\, \frac{\partial f_1}{\partial x} \nonumber
\end{eqnarray}

The integration over the radius $r$ is much simpler:
 \begin{subequations}
\label{eq:prima_r}
\begin{eqnarray}
\label{eq:int0} \int_0^{\infty}  \,e^{-r/l_f}\,dr&=& l_f \\
\label{eq:int1} \int_0^{\infty}r  \,e^{-r/l_f}\,dr&=& l_f^2 \\
\label{eq:int2} \int_0^{\infty}r^2 \,e^{-r/l_f}\,dr&=&2 l_f^3
\end{eqnarray}
\end{subequations}

The rate of momentum entering the volume element $\mathrm{d}V$ is then:
\begin{equation}
\label{eq:5termsApp}
\frac{\mathrm{d}p_\alpha}{\mathrm{d}t}  \simeq \mathrm{d}V  \, { \gamma _0}   \frac{B_v}{4}   \left(  \frac{3 T_P^4 v_{P\alpha}}{ l_f c } -4T_P^3   \, \frac{\partial T_P}{\partial \alpha}+8T_P^3    \frac{l_f}{ c } \, \nabla T_P\cdot\nabla  {v}_{P\alpha}+ 12 \frac{l_f}{c }\,T_P^2 \, {v}_{P\alpha}(\nabla T_P)^2 +  \frac{l_f}{c } T_P^4 \,\nabla^2{v}_{P\alpha}\right)
\end{equation} 

In respect to the I order expansion in $\textbf{r}$ leading to the Equation (7) of the main text, now all contributions up to the first order in the mean free path $l_f$ are considered. Further terms of the expansions \eqref{eq:vrII} and \eqref{eq:TSII} would add negligible contributions proportional to $l_f^2$. 
We have already discussed the first three elements of this equation, respectively the "thermal equilibrium", "phonon pressure" and "damping" term, and address now the relevance of the additional terms. 

\begin{itemize}
\item 
The fourth term is a correction to the damping term, as it depends both on the temperature gradient and on the local velocity of the crystal. In the case of a silicon cantilever discussed in the main text, the correction is less than 10\% over the considered range of relative gradient. We note that to derive this term we used the relation:
\begin{equation}
 \nabla^2 \left(T_P^4\right)= 
 3T_P^2(\nabla T_P)^2 +T_P^3 \nabla^2T_P  =3\,T_P^2(\nabla T_P)^2
\end{equation}
where  $\nabla^2 T_P =0 $ because $T_P$ is  a solution of the stationary heat equation with no source term.

\item
The fifth term is a correction to the thermal equilibrium term, therefore the associated loss angle concurs to determine the background loss. It  can augment or reduce the overall damping of the resonator, depending on the modal shape and on the Gr$\ddot{\textrm{u}}$neisen parameter $\gamma$ of the lattice. Lacking a the dependence from the gradient, we believe that it is not reasonable to single out this phononic contribution among the other dissipative mechanism.
As a consistency check, we only note that for the silicon cantilever discussed in the main text the expected contribution is a few units $\times 10^{-6}$, not far from the values measured in similar systems \cite{reid06}. This term is consistent also with the extremely low loss factor measured in resonators exploiting torsional modes, such as the  Double Paddle Oscillator \cite{spiel01}. In fact in this case the expected damping from phonon transport is orders of magnitude smaller due to the symmetry of the modal shapes.
\end{itemize}

\newpage
\section{Volume integral: medium temperature approximation for silicon}

From Equation \eqref{eq:sup_entering_dt_main0},  in the medium temperature case, we write the rate of momentum entering the volume element $dV$ as:
\begin{equation}
\label{eq:MTsup_entering_dt_main}
\frac{\mathrm{d}\textbf{p}}{\mathrm{d}t}=\mathrm{d}V\,    \frac{3   }{4 \pi  c \, l_f^2}  \int_{V_B} \mathrm{d} \sigma F_{MT}(T_Q)    \frac{1  }{  r^2  }\,e^{-r/l_f}  \left(-c\,\textbf{i}_r+\textbf{v}_Q  \right)  
\end{equation}
with $F_{MT}(T_Q)$ given by Equation \eqref{eq:FMT}.
In this case it is straightforward to show that, to the first order in $l_f$, the contributions to the rate of momentum entering the volume element $\mathrm{d}V$ are:
\begin{itemize}
\item Thermal equilibrium term
\begin{equation}
\mathrm{d}V\, \left( \frac{3 v_{p\alpha}}{l_f c}+ \frac{l_f}{c}\nabla^2 v_{p\alpha}\right)
  \left( F_0+F_1T_P+F_2T_P^2\right)
\end{equation}
\item Phonon pressure term
\begin{equation}
-\mathrm{d}V\, \frac{\partial T_P}{\partial \alpha}
  \left( F_1+2F_2T_P\right)
\end{equation}
\item Damping term
\begin{equation}
\mathrm{d}V\, \frac{2 l_f}{c}
  \left( F_1+2F_2 T_P\right)\nabla T_P\cdot\nabla  {v}_{P\alpha}+\mathrm{d}V\, \frac{2 l_f}{c} F_2 v_{p\alpha}(\nabla T_P)^2
\end{equation}

\end{itemize}
These local contributions were integrated over the time and over the volume of the cantilevered beam  to obtain the total loss angle at 100 K and 150 K reported in Figure 2 of the main text.